# Simulation of an Electrostatic Energy Harvester at Large Amplitude Narrow and Wide Band Vibrations

Lars Geir Whist Tvedt, Lars-Cyril Julin Blystad and Einar Halvorsen

Institute of Microsystem Technology
Faculty of Science and Engineering
Vestfold University College
P.O. Box 2243, N-3103 Tønsberg, Norway
E-Mail: Einar.Halvorsen@hive.no Phone: +47-33037725, Fax: +47-33031103

*Abstract* – **An electrostatic in-plane overlap varying energy harvester is modeled and simulated using a circuit simulator. Both linear and nonlinear models are investigated. The nonlinear model includes mechanical stoppers at the displacement extremes. Large amplitude excitation signals, both narrow and wide band, are used to emulate environmental vibrations. Nonlinear behavior is significant at large displacement due to the impact on mechanical stoppers. For a sinusoidal excitation the mechanical stoppers cause the output power to flatten and weakly decrease. For a wide band excitation, the output power first increases linearly with the power spectral density of the input signal, then grows slower than linearly.**

## I. INTRODUCTION

Decreasing power consumption of autonomous sensor systems makes it possible to obtain longer periods of system operation before maintenance is needed. By getting rid of replacement or manual recharging of batteries, maintenance free operation should in principle be possible. As the surrounding environment holds energy in various forms, it can be utilized to provide the system with the needed electrical power. One possible power source that has received considerable attention in recent years is energy harvesting from motion [1].

The possible transduction mechanisms that can be exploited for motion micro energy harvesting are conventionally categorized as piezoelectric, electromagnetic and electrostatic. Of these, the electrostatic principle is the one that relies on the most mature MEMS processes. These devices typically employ the same construction elements as used for MEMS accelerometers, i.e. a proof mass suspended in some beams, and variable capacitors based on one electrode set attached to the moving mass and one set of fixed counter electrodes attached elsewhere in the structure.

How to bias electrostatic harvesters is a challenge. The idea of providing a voltage source to the harvester seems to defeat its purpose, but makes sense if the source is part of the energy storage and power handling electronics associated with the harvester [2]. In that case the harvester needs to be kick-started at the beginning of its operational life.

A design that employs an electret as an internal bias and therefore avoids the need for external bias was proposed by Sterken et.al. [3]. A sketch of this type of electrostatic energy harvester with internal bias is shown in Fig. 1, top and middle drawing. In the equivalent circuit, bottom drawing of Fig. 1, the electret is represented by the voltage source $V_e$ in series with the capacitance $C_e$. An advantage of this design is that its transduction is quite insensitive to stray capacitances between the (large) proof mass and the package. This type of design is often referred to as an in-plane overlap converter [4]. The structure also can be used with an external bias.

The in-plane overlap varying converter was modeled in [3] by the use of equivalent circuit models. In [5], optimization of this device was studied with emphasis on the effect of nonlinearities for a fixed drive frequency and amplitude. Nonlinearities were further studied under sinusoidal excitation in [6]. All these analyses were made for sinusoidal excitations with amplitudes sufficiently small that the capacit-ance is still given by the plate capacitor formula and the effect of mechanical stoppers can be neglected.

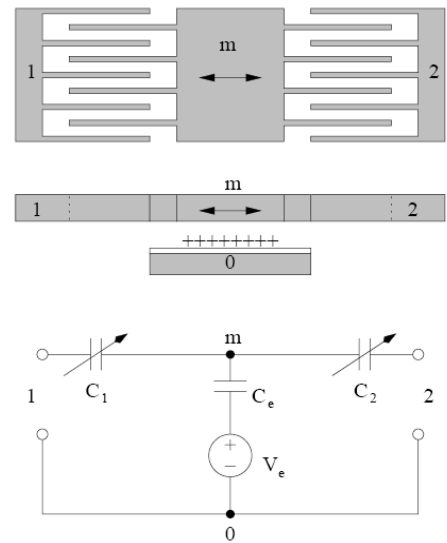

Fig. 1: Schematic view of an electrostatic energy harvester. The harvester has one ground electrode (0), a moving electrode (m), and two static electrodes (1 and 2). Top: Top view. Middle: Side view. Bottom: Equivalent circuit.

In the present work we investigate the effect of the acceleration magnitude on an in-plane overlap varying device by simulation. We consider the full range from the linear regime and well into the nonlinear regime where the driving signal is sufficiently large that the effect of mechanical stoppers must be considered. For simplicity we consider rigid stoppers and elastic impacts only. In addition to a sinusoidal driving signal, we also consider broadband excitations with a flat power spectral density (PSD) from zero frequency up to a frequency well above the harvester resonance frequency. The method is similar to that described in [7].

The simulations are made using a design that we have targeted for the current Tronics multi project wafer process using silicon on insulator (SOI) wafer technology with a device layer thickness of 60µm [8].

## II. Energy Harvester Model

In order to simulate the energy harvester performance, we have made a lumped element model. The model is based on a single mechanical degree of freedom (position of proof mass) and two electrical degrees of freedom (charge on the variable capacitors). A schematic model of the device in Fig. 1 is shown in Fig. 2 and consists of a moving mass, its suspension in the form of a spring, mechanical damping through a dashpot and two variable capacitors that vary out of phase as indicated. Two parasitic capacitances are also included.

The electromechanical transduction is described by the equations

$$F_T = kx + \frac{q_1^2}{2}\frac{d}{dx}\left(\frac{1}{C_1(x)}\right) + \frac{q_2^2}{2}\frac{d}{dx}\left(\frac{1}{C_2(x)}\right) \quad (1)$$

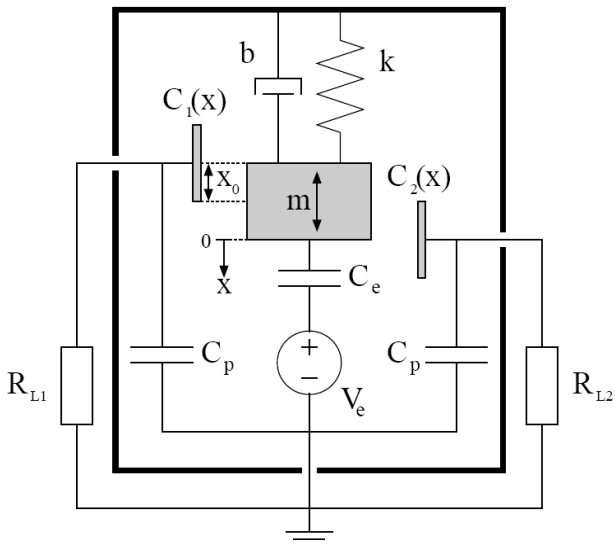

Fig. 2: Model of the energy harvester including both the mechanical and electrical subsystems.

$$V_{L1} = V_e + \frac{q_1+q_2}{C_e} + \frac{q_1}{C_1(x)} \quad (2)$$

$$V_{L2} = V_e + \frac{q_1+q_2}{C_e} + \frac{q_2}{C_2(x)} \quad (3)$$

where $F_T$ is the force exerted on the transducer, $V_{L1}$ and $V_{L2}$ are the voltages across the electrical ports, $x$ is the displacement of the proof mass, $k$ is the spring stiffness, $V_e$ is the voltage of the electret, $C_e$ is the capacitance of the electret, and $q_1$ ($q_2$) is the charge on the variable capacitor $C_1$ ($C_2$).

For the in-plane gap overlap converter, the capacitances of the variable capacitors are given by

$$C_1(x) = \frac{2 N_g \varepsilon\, t_f(x_0-x)}{g_0} \quad (4)$$

$$C_2(x) = \frac{2 N_g \varepsilon\, t_f(x_0+x)}{g_0} \quad (5)$$

where $N_g$ is the number of fingers of the comb structure, $\varepsilon$ is the permittivity of the medium between the two capacitor electrodes, $t_f$ is the thickness of the fingers (equal to the device layer of the SOI wafer), $x_0$ and $g_0$ are the nominal overlap of and gap between the finger structures of the capacitor electrodes respectively.

Linearization of (1-3) yields the following three equations for the electromechanical transduction

$$F_T = kx + \alpha_1 \Delta q_1 + \alpha_2 \Delta q_2 \quad (6)$$

$$V_{L1} = \alpha_1 x + \frac{\Delta q_1+\Delta q_2}{C_e} + \Delta q_1 \left.\frac{1}{C_1(x)}\right|_{x=0} \quad (7)$$

$$V_{L2} = \alpha_2 x + \frac{\Delta q_1+\Delta q_2}{C_e} + \Delta q_2 \left.\frac{1}{C_2(x)}\right|_{x=0} \quad (8)$$

where $\Delta q_1$ and $\Delta q_2$ are small changes in the charge on the two capacitor electrodes. $x$ is a small value itself since the equilibrium position is $x=0$. The couplings between the mechanical and electrical domains are given by $\alpha_1$ and $\alpha_2$:

$$\alpha_i = q_0 \left.\frac{d}{dx}\frac{1}{C_i(x)}\right|_{x=0} \quad i = 1,2. \quad (9)$$

where $q_0$ is the equilibrium charge on either capacitor.

In the lumped model, we include elastic stoppers to allow driving the system at large amplitude accelerations. In the model, they are represented by springs that are only in effect at sufficiently large displacements (Fig. 3).

The purpose of the stoppers is to limit the mass displace-ment, as to prevent structural damage, sticking, and avoid short circuiting of the electrodes. Stoppers must obviously be implemented in a real device, and should also be present in the model, as they affect the harvester operation at sufficiently large mass displacements. Stoppers are implemented in several previous designs [3, 9–11], but little has been reported on how stoppers are modeled.

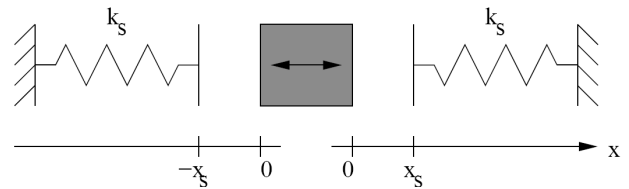

Fig. 3: Representation of elastic stopper model.

The elastic stoppers are modeled as opposing spring forces when in contact:

$$F_S = \begin{cases} 0 & , \; -x_s \leq x \leq x_s \\ -k_s(x + x_s), & x < -x_s \\ -k_s(x - x_s), & x > x_s \end{cases} \quad (10)$$

where $F_s$ is the force of the stoppers, $k_s$ is the spring stiffness of the stoppers and $x_s$ is the displacement at which the stoppers starts acting (maximum possible displacement of the energy harvester). We choose $k_s$ very large compared to $k$ to mimic real hard walls.

To avoid unreasonable capacitance variations at large displacements, the variable capacitances are "cut off" at both ends as in Fig. 4 and can be expressed as

$$C_1(x) = \begin{cases} \frac{2N_g \epsilon t_f (x_0 - x)}{g_0} & , \; -x_c \leq x \leq x_c \\ C_{\max} & , \; x < -x_c \\ C_{\min} & , \; x > x_c \end{cases} \quad (11)$$

$$C_2(x) = \begin{cases} \frac{2N_g \epsilon t_f (x_0 + x)}{g_0} & , \; -x_c \leq x \leq x_c \\ C_{\min} & , \; x < -x_c \\ C_{\max} & , \; x > x_c \end{cases} \quad (12)$$

where $C_{max}$ and $C_{min}$ are the limiting values of the variable capacitances. $x_c$ is the displacement of the mass at which the limits are reached. The cut-off on the minimum capacitance is necessary to avoid an unphysical negative value when the overlap goes through zero. We have $x_c > x_s$ so that the cut-off is rarely in effect.

Motion of the mass inside the cavity will create gas flow along the mass and its finger structures. We assume Couette flow, giving the damping coefficient

$$b = 2\eta \left( \frac{N_g t_f l_f}{g_0} + \frac{A_m}{d} \right) \quad (13)$$

where $\eta$ is the viscosity of the gas, $l_f$ is the length of the capacitor fingers, $A_m$ is the area of the mass and $d$ is the distance between the mass and the top and bottom cap of the cavity in which the mass is moving.

Newton's second law for the mass gives

$$m\ddot{x} = -F_T - F_S - b\,\dot{x} + ma \quad (14)$$

where $m$ is the mass and $ma$ is the fictitious force experienced due to the acceleration of the device.

For the electrical degrees of freedom, the equations of motion depend on the load represented by the resistors $R_{L1}$ and $R_{L2}$ and stray capacitances in Fig. 2. Since the model is implemented in a circuit simulator in section III, it can be simulated with almost arbitrary power conditioning circuitry, but we have chosen to use simple load resistors to keep focus on the characteristics of the MEMS device.

The dimensions of the energy harvester are given in Table I and derived model parameters are given in Table II.

### III.  Spice Implementation

The model is represented by an equivalent circuit using the e-V convention which we have implemented in a SPICE simulator. For the mechanical part, the forces are represented by voltages, displacements by charges and velocities by currents. Stiffness is then represented by a capacitor of "capacitance" $1/k$, mechanical damping by a resistor of "resistance" $b$, and inertia by an inductor of "inductance" $m$.

The sinusoidal vibration signal is represented as a sinusoidal voltage source of a given frequency and amplitude. For broadband vibrations a voltage source with input from a source file with a synthetically generated random broadband signal are used.

The mechanical stoppers are implemented using a behavioral voltage source.

TABLE I
DIMENSIONS OF THE ENERGY HARVESTER

| | |
|---|---|
| Length of capacitor fingers, $l_f$ | 30 μm |
| Width of capacitor fingers, $w_f$ | 4 μm |
| Thickness of capacitor fingers, $t_f$ | 60 μm |
| Gap between capacitor fingers, $g_0$ | 3 μm |
| Number of capacitor finger pairs, $N_g$ | 524 |

TABLE II
MODEL PARAMETERS OF THE ENERGY HARVESTER

| | |
|---|---|
| Mass, $m$ | 5.78 mg |
| Spring constant, $k$ | 326 N/m |
| Damping constant, $b$ | 8.45×10⁻⁴ N-s/m |
| Electret voltage, $V_e$ | 20 V |
| Electret capacitance, $C_e$ | 5 pF |
| Parasitic capacitance, $C_p$ | 1,94 pF |
| Initial finger overlap, $X_0$ | 15 μm |
| Displacement limit of stoppers, $X_s$ | 14 μm |
| Spring constant of stoppers, $k_s$ | 326×10³ N/m |
| Displacement limit of variable capacitors, $X_c$ | 14.5 μm |

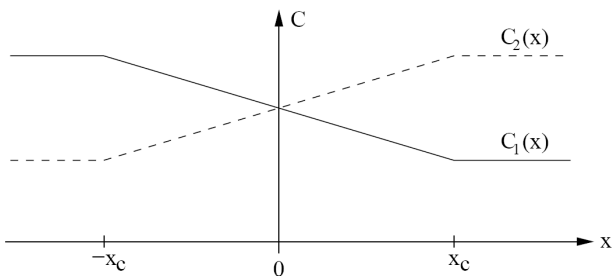

Fig. 4: Representation of the limitations done to the variable capacitances and their derivatives.

## A. Linear SPICE Model

Fig. 5 shows the linear lumped model equivalent circuit based on (6-8) and used for the SPICE simulations. We have used behavioral voltage sources to implement the coupling between the mechanical and electrical parts of the model. In the behavioral model, $x$ is found by using the voltage $V_m$ across the capacitor and its value.

The main purpose of the linear model is to compare to the more realistic nonlinear model and identify their differences. We therefore do not include stoppers in the linear model.

$C_0$ is the nominal capacitance of the variable capacitors, and $C_P$ models the parasitic capacitances at the output nodes.

## B. Nonlinear SPICE Model

Fig. 6 shows the nonlinear lumped model equivalent circuit based on (1-3) used for the SPICE simulations. The variable capacitors, the last element in (2) and in (3), are realized using a behavioral voltage source in series with a fixed capacitor with value $C_0$. $V_0$ is the voltage across the fixed capacitor. The voltage, $V_S$, is then given by:

$$V_{S1} = V_0 \left( \frac{C_0}{c_1(x)} - 1 \right) \quad (15)$$

$$V_{S2} = V_0 \left( \frac{C_0}{c_2(x)} - 1 \right) \quad (16)$$

## IV. SIMULATION RESULTS AND DISCUSSION

To investigate the effect of increasing excitation strength both for sinusoidal and for wide band excitation, we first found the optimum load in the linear regime for these two cases. Using this as a starting point, we have performed simulations at various excitation levels.

Results from the simulations on the linear model for a sinusoidal input signal and with increasing frequency and load resistance are shown in Fig. 7. An optimum driving frequency of 1190 Hz and a load resistance of 28 MΩ have been found. In Fig. 8, for the same model with a broadband input signal and increasing load resistance, the same optimum load resistance of 28 MΩ was found. In general they will differ [12]. Due to the small electromechanical coupling they coincide here.

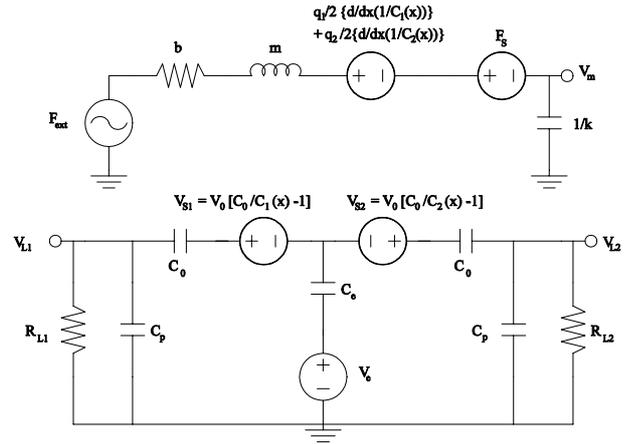

Fig. 6: Equivalent circuit used for simulations on the nonlinear energy harvester model in SPICE. Top: The mechanical part. Bottom: The electrical part.

Fig. 9 shows a close-up of phase space plots for the proof mass when driving with a sinusoidal signal. In the simulations the stopper is set to act at displacements larger than 14 μm. The action of the stoppers at displacements larger than this is visible at both 5 g and 10 g. At 3.7 g the displacements are at the limit of the maximum displacement before the stoppers begin to act. For small amplitude vibrations we have an elliptic looking trajectory, which becomes distorted as the stopper comes into effect.

Fig. 10 shows a phase space plot for the same model driven by a broadband random acceleration with a PSD of 0.015 g²/Hz. Due to the random character of the motion, the trajectory is not cyclic. Nevertheless, since the harvester is quite narrow band it has some cyclic like behavior. When the excitation is increased to 0.045 g²/Hz, we get the trajectory shown in Fig. 11. Here the effect of stoppers is clearly seen.

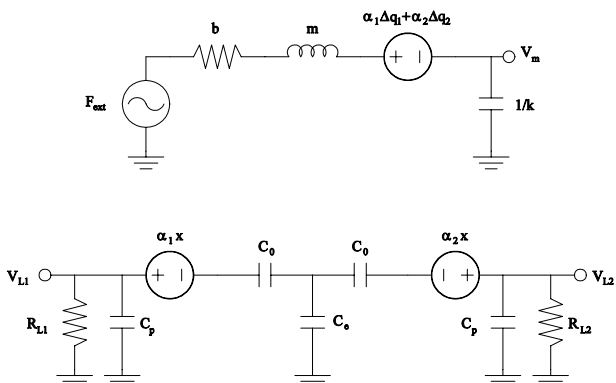

Fig. 5: Equivalent circuit used for simulations on the linear energy harvester model in SPICE. Top: The mechanical part. Bottom: The electrical part.

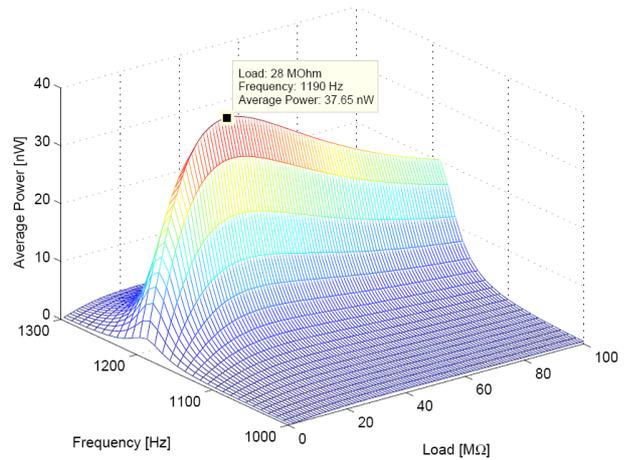

Fig. 7: Linear model, average power versus driving frequency and load resistance at a sinusoidal input signal of 1 g acceleration. Filled square: Optimum power output at 1190 Hz and a load resistance of 28 MΩ.

                                    

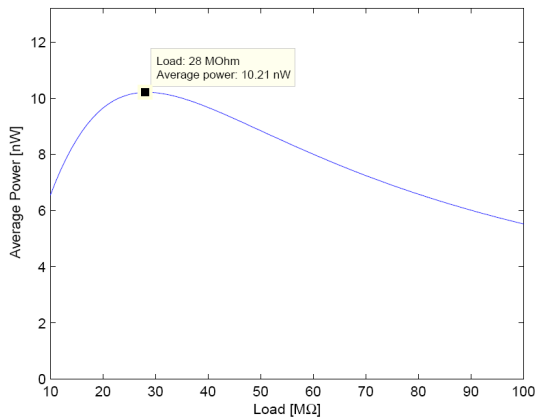

Fig. 8: Linear model, average power versus load resistance for a wideband input signal. Optimum output power is found with a load resistance of 28 MΩ.

Fig. 12 shows the output power of the linear versus the nonlinear model under sinusoidal excitation of fixed frequency. For the nonlinear model at low nonlinearity (before stoppers come into effect) the transduction is more effective and gives a larger output power than what is the case for the linear model. At larger displacements, where the stoppers act more frequently, the output of the nonlinear model flattens and starts to decrease weakly at increasing input vibration amplitude. As the stoppers come into effect, the frequency at which the motion is in phase with the excitation will start to increase; therefore transduction is much less effective at the fixed drive frequency which was optimal in the linear regime.

Fig. 13 shows the average output power and mean square displacement of the mass versus PSD of the input wideband signal for the nonlinear model. The output power increases linearly with the PSD for low level excitations, as it should according to theory [12]. For larger excitations the output power continues to increase, but at a lower rate. The reason why the stoppers have considerably less dramatic effect in the broad band case is that even if the harvester's sensitive frequency drifts as a function of drive level, there are always frequency components at this frequency in the excitation signal. This is true until the sensitive frequency drifts out of the excitation bandwidth.

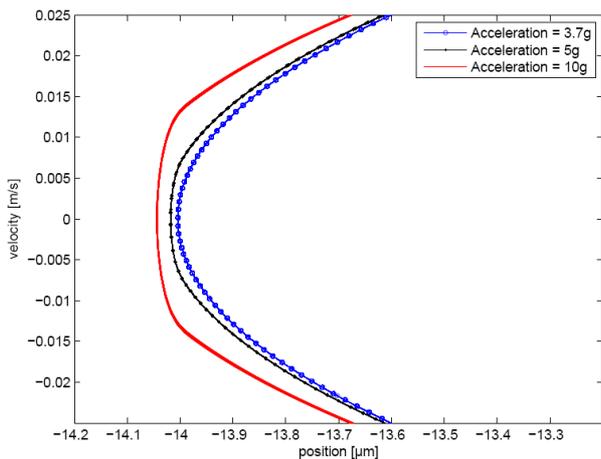

Fig. 9: Phase space plots for nonlinear model with sinusoidal excitations.

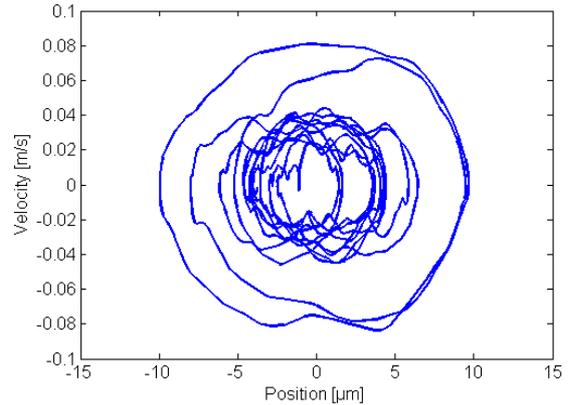

Fig. 10: Phase space plot for nonlinear model at wideband excitation with a PSD of 0.015 g²/Hz

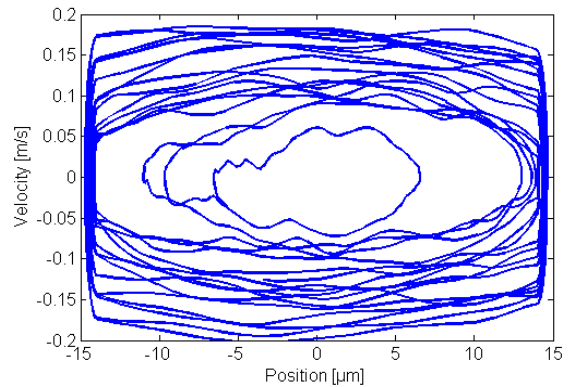

Fig. 11: Phase space plot for nonlinear model at wideband excitation with a PSD of 0.045 g²/Hz

## V.    CONCLUSIONS

We have made a lumped model of an electrostatic in-plane overlap varying energy harvester. The model includes both nonlinearities in the transduction mechanism and the effect of mechanical stoppers for large displacements.

We have demonstrated simulations of the energy harvester for both narrow and wide band excitations using a SPICE implementation of the lumped model.

For a sinusoidal input signal the nonlinear model has a more effective transduction than the linear model at low nonlinearity, but increasing the input signal strength beyond the point where the mechanical stoppers starts acting causes the output power to flatten and weakly decrease.

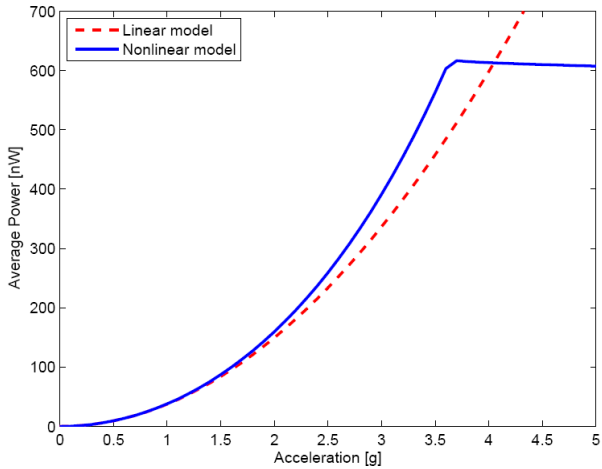

Fig. 12: Linear model versus nonlinear model. The nonlinear model has a more effective transduction than the linear model at low nonlinearity.

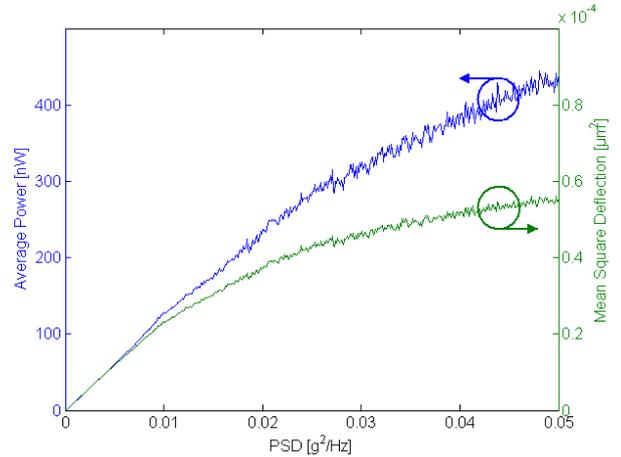

Fig. 13: Nonlinear model. Left: Average output power versus PSD. Right: Mean square mass displacement versus PSD.

For a wide band excitation, the output power increases linearly with the PSD of the input signal for low signal strength. As the excitation strength increases, the output power continues to increase, but grows slower than linearly with respect to the power spectral density. At a PSD of 0.05 g$^2$/Hz the output power is approximately 400nW.

ACKNOWLEDGMENT

We thank E. Westby and S. Husa for useful discussions.